\documentclass[lettersize,journal]{IEEEtran}

\usepackage{cite}

\usepackage{stfloats}

\usepackage{amsmath,amsfonts}
\usepackage[ruled,boxed,linesnumbered]{algorithm2e}

\usepackage{listings}
\lstdefinestyle{pseudocode}{
	basicstyle=\ttfamily\footnotesize,
	keywordstyle=\bfseries,
	morekeywords={Input,Output,for,each,in,do,if,then,else,return,FAIL,<-,func,continue,break,forall},
	numbers=none,
	numberstyle=\tiny,
	numbersep=6pt,
	frame=single,
	breaklines=true,
	columns=fullflexible,
	keepspaces=true,
	showstringspaces=false,
	captionpos=t
}

\usepackage{graphicx}
\usepackage{subfigure}

\usepackage{multirow}
\usepackage{makecell}
\usepackage{threeparttable}

\usepackage{stfloats}

\usepackage{textcomp}
\usepackage{xcolor}

\usepackage{verbatim}

\usepackage[colorlinks=true,linkcolor=blue,citecolor=blue,urlcolor=blue]{hyperref}

\usepackage{tikz}
\usepackage{tikz-timing}

\def\BibTeX{{\rm B\kern-.05em{\sc i\kern-.025em b}\kern-.08em
		T\kern-.1667em\lower.7ex\hbox{E}\kern-.125emX}}
	
\begin{document}



\title{TPCS: A Cycle-Level Simulation Paradigm with Native Execution Semantics for RTL Modeling and Generation}

\author{
	\IEEEauthorblockN{
		Chengzhen Meng\IEEEauthorrefmark{1},
		Xiaoyu Li\IEEEauthorrefmark{2},
		Chen Bai\IEEEauthorrefmark{3},
		Zhenyu Zhang\IEEEauthorrefmark{2},
		Hongjun Dai\IEEEauthorrefmark{1}\thanks{Corresponding author: Hongjun Dai (email: dahogn@sdu.edu.cn)}
	}
	
	\IEEEauthorblockA{\IEEEauthorrefmark{1} School of Software, Shandong University, Jinan, China}
	\IEEEauthorblockA{\IEEEauthorrefmark{2} School of Integrated Circuits, Shandong University, Jinan, China}
	\IEEEauthorblockA{\IEEEauthorrefmark{3} Fudan University, Shanghai, China}
	\thanks{This work has been submitted to the IEEE for possible publication. Copyright may be transferred without notice, after which this version may no longer be accessible.}
}

\maketitle

\begin{abstract}
	
	The growing diversity of domain-specific architectures and microarchitectural designs demands fast and accurate evaluation methods during early design stages.
	However, existing approaches suffer from a fundamental trade-off: high-level simulators offer speed but lack fidelity, while RTL-based tools provide accuracy at the cost of significant simulation overhead.
	This gap limits efficient pre-RTL microarchitecture exploration.
	Therefore, we propose Transaction-Pruned Cycle Simulation (TPCS), a novel modeling and simulation framework.
	It serves as a unified intermediate abstraction spanning microarchitecture modeling, cycle-accurate simulation, and synthesizable RTL implementation.
	TPCS organizes combinational logic as transaction-based function invocations and leverages host CPU control flow to naturally prune inactive logic, eliminating the need for explicit event scheduling or signal tracking.
	This enables cycle-accurate simulation with significantly improved performance.
	TPCS preserves structural correspondence with RTL and supports automatic generation of synthesizable designs.
	TPCS achieves 36$\times$-42$\times$ speedups over parallel Verilator on SMP systems and 3.4$\times$-7.2$\times$ over single-threaded Verilator on accelerators, with cycle-level waveform agreement on the evaluated traces.
	The generated RTL demonstrates comparable timing and area to handcrafted implementations, with relative changes of -4.7\% to +2.6\% in clock period and -3.3\% to +4.0\% in standard-cell area.
	By bridging the abstraction gap between high-level modeling and RTL design, TPCS enables a unified and agile pre-RTL workflow, allowing efficient microarchitecture exploration with both high performance and implementation fidelity.
	
\end{abstract}

\begin{IEEEkeywords}
Microarchitecture Modeling, Cycle-Accurate Simulation, Hardware/Software Co-Design, Design Methodology.
\end{IEEEkeywords}

\section{Introduction}

	Modern microarchitecture ($\mu$arch) innovation extends beyond conventional IP integration; fine-grained design has become equally critical, particularly in domain-specific processors and heterogeneous systems, where detailed $\mu$arch decisions significantly impact performance \cite{ref-dac-arch-2-0, ref-byg}.
	Rapid exploration of such designs requires efficient evaluation of $\mu$arch ideas before committing to error-prone and time-consuming Register-Transfer Level (RTL) implementation.
	Skipping this pre-RTL evaluation leads to high development costs, including long synthesis-simulation cycles, limited design-space coverage, and costly rework when sub-optimal decisions are discovered late \cite{ref-dac-uarch-explo, ref-dac-cpu-explo, ref-ripes}.
	Therefore, agile $\mu$arch iteration demands evaluation methods that are both fast for early exploration and sufficiently faithful to capture fine-grained hardware behavior.

\begin{figure} [t]
	\centering
	\subfigure[Lack of fine-grained $\mu$arch evaluation tools in conventional workflows.]{
		\begin{minipage}{\columnwidth}
			\centering
			\includegraphics[]{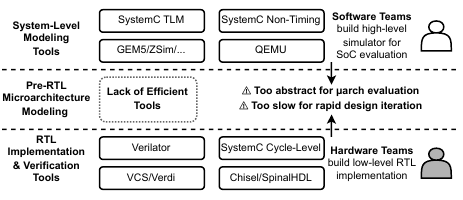}
		\end{minipage}
		\label{fig-intro1}
	}
	\subfigure[Proposed workflow with TPCS enabling agile pre-RTL iteration.]{
		\begin{minipage}{\columnwidth}
			\centering
			\includegraphics[]{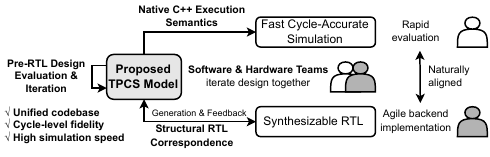}
		\end{minipage}
		\label{fig-intro2}
	}
	\caption{A typical workflow in fine-grained $\mu$arch design. TPCS addresses the bottleneck caused by abstraction gap and models/tools isolation.}
	\label{fig-intro}
\end{figure}
	
	However, existing simulation or evaluation tools fall into two opposing abstraction levels (Fig. \ref{fig-intro1}).
	This polarization stems from a fundamental mismatch in design philosophies: high-level tools (e.g., GEM5, TLM) prioritize exploration agility over detailed hardware behavior, while RTL tools and hardware description languages (HDLs) prioritize low-level fidelity but incur substantial simulation overhead.
	As a result, neither supports efficient and accurate pre-RTL $\mu$arch evaluation.
	This mismatch forces prior approaches to trade off between efficiency and fidelity \cite{ref-boom-sim, ref-eventsim-knightsim, ref-rvvec-sim-bench-nogem5, ref-survey-cachesim} or rely on complex multi-level workflows \cite{ref-gem5-rtl, ref-gem5-accelerator-extension, ref-gem5-accel, ref-gem5rtl-chimera}.
	Consequently, the resulting design misalignment, time overhead, and engineering effort form key bottlenecks in agile $\mu$arch iteration.
	
	Addressing these challenges is fundamentally difficult, as they originate from the absence of an execution model bridging high-level programmability and low-level hardware behavior.
	This motivates a rethinking of how hardware execution is represented and simulated.
	
	The overhead of RTL simulation largely stems from signal-level evaluation, where even inactive computations incur unnecessary work.
	Prior studies reduce this overhead by tracking signal activity \cite{ref-gsim, ref-rtl-act}, but they remain confined to the signal-level abstraction.
	We observe that hardware behavior is inherently governed by control logic, which determines the activation of entire computation blocks.
	Based on this, we reformulate hardware simulation as an execution process based on behavioral descriptions.
	Inactive paths are naturally skipped via host-native control flow, eliminating the need for explicit activity tracking.
	This formulation not only improves simulation efficiency, but also provides an executable representation of hardware behavior prior to RTL realization.
	
	In this paper, we propose Transaction-Pruned Cycle Simulation (TPCS), a novel methodology for modeling and simulating deterministic cycle-synchronous hardware designs using native software descriptions.
	TPCS represents hardware execution as transaction-level function invocations within each cycle, enabling control-flow-driven execution with implicit pruning of inactive logic.
	This design achieves three key properties (Fig. \ref{fig-intro2}):
	(1) cycle-level equivalence with RTL at module boundaries,
	(2) high simulation efficiency through native execution, and
	(3) structural correspondence that supports automatic generation of synthesizable RTL.
	Together, these properties enable fast and accurate pre-RTL design evaluation for agile hardware development.

	We evaluate TPCS across designs with varying control intensity.
	TPCS achieves 36$\times$--42$\times$ speedups on SMP systems and 3.4$\times$--7.2$\times$ on accelerators against parallel and single-threaded Verilator baselines, respectively.
	Meanwhile, waveform equivalence is preserved for key registers and module boundaries.
	In addition, the generated RTL achieves timing and area close to hand-written designs, with relative changes of -4.7\% to +2.6\% in clock period and -3.3\% to +4.0\% in standard-cell area.
	These results demonstrate that TPCS delivers both high simulation performance and implementation-quality hardware.
	
	The main contributions of this paper are as follows:
	\begin{itemize}
		\item We introduce a new hardware simulation paradigm that leverages host-native control flow to directly drive the control logic of the simulated target hardware.
		\item We formalize this idea into TPCS, a novel hardware modeling method that combines native host-language execution, transaction-driven pruning, and systematic generation of synthesizable RTL. The toolchain implementing TPCS is open-source \footnote{https://github.com/meng-cz/vulsim}.
		\item We demonstrate TPCS on representative cycle-synchronous hardware designs, showing cycle-level waveform equivalence, up to 42$\times$ speedup over parallel Verilator, and generated-RTL quality comparable to hand-written designs.
		\item We show that TPCS enables an agile pre-RTL development workflow, allowing design issues to be identified and resolved earlier, thereby reducing iteration time and engineering effort compared to conventional workflows.
	\end{itemize}
	
\section{Related Work}

\begin{table}[]
	\centering
	\caption{Comparison of TPCS and Typical Related Works.}
	\resizebox{\linewidth}{!}{
	\begin{threeparttable}
	\begin{tabular}{|c|c|c|c|c|c|c|c|c|}
		\hline
		\textbf{Work} & \textbf{ABS\textsuperscript{1}} & \textbf{GEN\textsuperscript{2}} & \textbf{ACC\textsuperscript{3}} & \textbf{SPD\textsuperscript{4}} & \textbf{SIM\textsuperscript{5}} & \textbf{EXE\textsuperscript{6}} & \textbf{PAR\textsuperscript{7}} & \textbf{RG\textsuperscript{8}} \\ \hline
		Verilator & RTL  & ALL  & CYC & $\star\star$ & CYC & TRA & YES & N/A \\
		VCS       & RTL  & ALL  & GAT & $\star$      & EVT & TRA & YES & N/A \\
		GSIM\cite{ref-gsim}      & RTL  & ALL  & CYC & $\star\star\star$ & CYC & TRA & YES & N/A \\
		MetroMPI\cite{ref-rtl-sim-mpi}  & RTL  & ALL  & CYC & $\star\star\star$ & MIX & TRA & YES & N/A \\ \hline
		SystemC-TLM\cite{ref-sysmtec-tlm}    & TLM  & PROC & APX & $\star\star\star\star$ & EVT & SCH & NO & NO \\
		GEM5\cite{ref-gem5}      & $\mu$ARCH-APX & PROC & APX & $\star\star\star\star$ & EVT & SCH & NO & NO \\
		GEM5+RTL\cite{ref-gem5-rtl}  & $\mu$ARCH-APX & PROC & APX & $\star\star\star$ & MIX & TRA & NO & NO \\ \hline
		SystemC-Cycle  & RTL  & ALL  & CYC & $\star\star$ & CYC & SCH & NO & YES \\
		Chisel\cite{ref-chisel}    & RTL  & ALL  & CYC & $\star$      & CYC & TRA & NO & YES \\
		PyMTL\cite{ref-pymtl}     & RTL  & ALL  & CYC & $\star$      & CYC & TRA & NO & YES \\ \hline
		Assassyn\cite{ref-assassyn}  & $\mu$ARCH-CYC & PIPE & CYC & $\star\star$ & CYC & TRA & NO & YES \\
		\textbf{TPCS} & $\mu$ARCH-CYC & GEN & CYC & $\star\star\star\star$ & CYC & NAT & YES & YES \\ \hline
	\end{tabular}
	\begin{tablenotes}
	\footnotesize
	\item[1] Abstraction Level, RTL: Register-transfer level, $\mu$ARCH-APX: Latency-based $\mu$arch model, $\mu$ARCH-CYC: Cycle-defined $\mu$arch model, TLM: Transaction-level model.
	\item[2] Hardware Generality, GEN: General cycle-synchronous hardware, PIPE: Pipeline-based hardware, PROC: Processor \& SoC only.
	\item[3] Timing Accuracy, GAT: Gate-level, CYC: Cycle-level, APX: Approximate timing.
	\item[4] Simulation Speed.
	\item[5] Simulation Model, EVT: Event-driven, CYC: Cycle-driven, MIX: Mixed.
	\item[6] Executability, NAT: Natively executable with static scheduling, SCH: Dynamic scheduler required, TRA: Translation required before execution.
	\item[7] Parallel Simulation.
	\item[8] RTL Generation.
	\end{tablenotes}
	\end{threeparttable}
	}
	\label{tab-relatedwork}
\end{table}

	Prior work has extensively explored accelerating software-based RTL simulation through parallelization and signal-update pruning techniques \cite{ref-gsim, ref-rtl-act, ref-rtl-sim-mpi, ref-rtl-khronos, ref-rtl-loop-sim, ref-soft-rtl-sim-hls}.
	Representative efforts include GSIM \cite{ref-gsim}, which is among the fastest reported RTL simulators with multi-level optimizations, and MetroMPI \cite{ref-rtl-sim-mpi}, which scales to large designs via distributed simulation by scheduling events across nodes using MPI, thereby exploiting parallelism at scale.
	However, RTL-based simulation faces a fundamental limitation: the mismatch between RTL's signal-driven semantics and the host CPU's control-flow execution requires explicit scheduling or control logic, introducing inherent overhead.
	In contrast, TPCS is designed to map hardware behavior natively onto the host instruction stream, allowing control flow to be handled implicitly and eliminating this overhead.
	Verilator is used as the RTL simulation baseline due to its widespread adoption, robustness across general hardware designs, and consistent reproducibility, making it a reliable point of comparison despite the existence of newer, less mature alternatives.
	
	Mixed frameworks improve simulation efficiency by combining different modeling abstractions and selectively omitting low-level hardware details \cite{ref-sysmtec-tlm, ref-tlm-cpu, ref-gem5-rtl, ref-gem5rtl-chimera, ref-gem5-hetosoc-sim, ref-pyrtl-cocotb}.
	Representative efforts include SystemC TLM \cite{ref-sysmtec-tlm} with multiple timing models and GEM5+RTL \cite{ref-gem5-rtl} integrating RTL components into an architectural simulator.
	While these approaches achieve higher performance with limited loss of accuracy, they do not fundamentally bridge abstraction levels.
	Hardware must still be re-implemented across layers, and additional interface overhead is unavoidable.
	In contrast, TPCS focuses on a unified intermediate representation of hardware rather than cross-level composition.

	Cycle-level synthesizable SystemC can describe hardware at RTL-like granularity and support synthesis, but its execution semantics remain mediated by the SystemC kernel through process scheduling and signal/event updates.
	TPCS targets a different point in this design space: it retains cycle-level state and structural RTL correspondence, while mapping hardware control flow directly onto native host-language function calls and control flow, without a runtime event scheduler.

	High-level RTL generation has become a key direction for agile hardware design, but it typically relies on well-explored architecture and parameterized generators, limiting its applicability to specific domains such as CPUs \cite{ref-riscv-gen, ref-cpugen-pdl, ref-dac-llm-cpu-gen, ref-chipyard}, accelerators \cite{ref-gem5-accel, ref-end2end-sim-dlworkload, ref-comp-driv-sim}, and SoCs \cite{ref-chipyard, ref-hetosim-fullstack}.
	More generally, high-level hardware design frameworks focus on generating RTL from software-like intermediate representations such as Chisel \cite{ref-chisel, ref-firrtl, ref-chiselfv, ref-pymtl}, improving design productivity but remaining signal-level construction.
	
	The work most closely related to TPCS is Assassyn \cite{ref-assassyn}, which introduces a unified "C-like" pipeline stage representation and can generate both cycle-accurate simulation and synthesizable RTL.
	However, it is limited to regular hardware structures based on pipelines and FIFOs, and its event-driven, single-threaded simulation model constrains performance.
	In contrast, TPCS adopts a native C++ representation and leverages transaction-based, control-flow-driven execution semantics, achieving broader hardware generality and simulation speed close to software simulators.
	\textbf{TPCS represents a step toward reconciling software execution with cycle-accurate hardware semantics for modeling and simulating general synchronous hardware}, enabling a more agile hardware design workflow.

\section{Transaction-Pruned Cycle Simulation}
	
\subsection{Overview of TPCS}
	
	\begin{figure} [t]
		\centering
		\subfigure[Behavioral abstraction: a TPCS modeled producer-consumer example.]{
			\begin{minipage}{\columnwidth}
				\centering
				\includegraphics[]{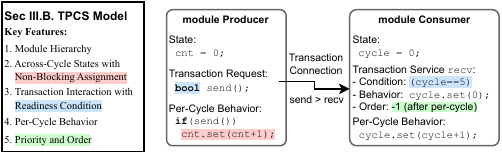}
			\end{minipage}
			\label{fig-overview1}
		}
		\subfigure[Software-level: Executable C++ code assembled from the TPCS model.]{
			\begin{minipage}{\columnwidth}
				\centering
				\includegraphics[]{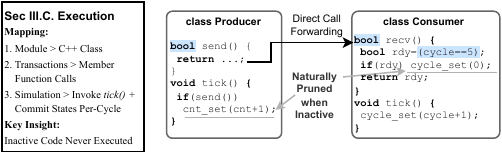}
			\end{minipage}
			\label{fig-overview2}
		}
		\subfigure[Hardware-level: Synthesizable RTL generated from the TPCS model.]{
			\begin{minipage}{\columnwidth}
				\centering
				\includegraphics[]{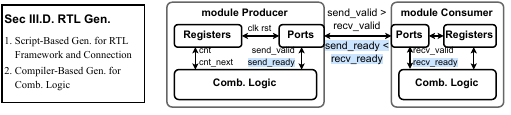}
			\end{minipage}
			\label{fig-overview3}
		}
		\caption{Overview of TPCS modeling, execution, and RTL generation.}
		\label{fig-overview}
	\end{figure}
	
	To bridge high-level modeling and RTL fidelity, we propose Transaction-Pruned Cycle Simulation (TPCS), a modeling and simulation paradigm with native execution semantics. 
	The term “Transaction-Pruned” refers to the elimination of inactive code execution in each cycle via transaction-based control-flow conditions, thereby pruning unnecessary simulation work.
	
	TPCS comprises three aspects: (1) modeling abstraction (Section \ref{sec-method-model}), (2) execution semantics (Section \ref{sec-method-exec}), and (3) RTL generation (Section \ref{sec-method-rtl}). 
	Fig. \ref{fig-overview} illustrates the TPCS workflow using a producer-consumer example, which will be referenced throughout this section.
	
	TPCS differs from traditional HDLs by focusing on behavior rather than construction (Fig. \ref{fig-overview1}). 
	The key challenge is to give a precise and executable definition of \textit{behavior}. 
	To this end, TPCS defines each hardware module in three aspects: 
	(1) A set of states representing cycle-boundary observable values, with non-blocking assignment effective at next cycle (pink-shaded texts).
	(2) Transactions capturing inter-module control and data interactions via request (caller) and service (callee), with transaction readiness condition (blue-shaded texts).
	(3) Behavioral procedures expressed as C++ code operating on states and transactions within a cycle. 
	Together, these define the complete per-cycle semantics of a module as deterministic state transitions and interactions under given inputs. 
	This unified abstraction forms a precise and executable pre-RTL representation that is both human-readable and structurally aligned with RTL.
	
	At the software level, TPCS models are directly executable: advancing one cycle corresponds to executing per-cycle behaviors of all modules (Fig. \ref{fig-overview2}). 
	The key challenge is to achieve high performance while preserving cycle accuracy. 
	TPCS addresses this by statically resolving control flow.
	State assignments are mapped to C++ APIs, and transaction interactions are mapped to function calls.
	Therefore, the execution of a cycle is driven entirely by host-language semantics without global schedulers or event queues.
	Transaction conditions and control-flow branches naturally prune inactive code at runtime, incurring only minimal branch prediction cost and no additional activity tracking. 
	As a result, TPCS achieves near high-level simulation performance while maintaining cycle-level fidelity.
	
	At the hardware level, the structured abstraction of TPCS enables systematic generation of synthesizable RTL (Fig. \ref{fig-overview3}). 
	The key challenge is to translate behavior and transaction into signal-level implementations while preserving semantics. 
	TPCS adopts a structure-logic decoupled approach.
	Module hierarchy, state, and transactions are directly mapped to an RTL skeleton with modules, registers, and port connections.
	Behavioral procedures are translated into equivalent combinational logic via the dedicated TPCS compiler. 
	This separation establishes a consistent mapping from high-level behavior to signal-level implementation, ensuring equivalence in state transitions and interactions between the TPCS model and generated RTL.

\subsection{TPCS Modeling Semantics with RTL Correspondence}
\label{sec-method-model}

	\begin{table*}[]
		\centering
		\caption{TPCS, C++, and RTL element mapping.}
		\resizebox{0.95\linewidth}{!}{
		\begin{tabular}{|c|l|l|l|}
			\hline
			\textbf{Category} & \textbf{TPCS Element} & \textbf{C++ Element} & \textbf{RTL Element} \\ \hline
			\multirow{2}{*}{Declaration} & \texttt{\textbf{CONSTANT}(Num, 16)} & \texttt{\textbf{constexpr int64} Num = 16;} & \texttt{\textbf{localparam longint} Num = 16;} \\ \cline{2-4}
			& \makecell[l]{\texttt{\textbf{STRUCT}(Msg) \{} \\ \texttt{ \textbf{uint8} data;} \\ \texttt{\}}} & \makecell[l]{\texttt{\textbf{struct} Msg \{} \\\texttt{ \textbf{uint8} data;} \\  \texttt{\}} } & \makecell[l]{\texttt{\textbf{struct packed} \{} \\ \texttt{ \textbf{logic} [7:0] data;} \\ \texttt{\} Msg;} }  \\ \hline
			\multirow{4}{*}{\makecell[c]{Module \\ Definition \\ (Sec. \ref{sec-method-module}) } } & \makecell[l]{\texttt{\textbf{MODULE}(Sender)} \\ \texttt{\textbf{PARAMETER}(WIDTH, 4)}} & \makecell[l]{\texttt{\textbf{template}<\textbf{int64} WIDTH = 4>} \\ \texttt{\textbf{class} Sender \{ ... \};} } & \makecell[l]{\texttt{\textbf{module} Sender \#(} \\ \texttt{ \textbf{parameter longint} WIDTH = 4);} } \\ \cline{2-4}
			& \texttt{\textbf{INSTANCE}(Sender, send, WIDTH=8)} & \texttt{\textbf{unique\_ptr}<Sender<8>> send;} & \texttt{Sender \#(.WIDTH(8)) send (...); } \\ \cline{2-4}
			& \makecell[l]{\texttt{\textbf{PERSISTENT\_STATE}(Msg, buf, 2)} \\ \texttt{// type, name, write port number} } & \makecell[l]{ \texttt{PState<Msg, 2> buf;} \\ \texttt{\textbf{void} buf\_set(\textbf{const} Msg\textbf{\&} v,} \\ \texttt{\textbf{\ \ const uint32} prio = 0) \{ } \\ \texttt{ buf.set(v, prio);\} } } & \makecell[l]{ \texttt{\textbf{reg} [7:0] buf;} \\ \texttt{\textbf{wire} buf\_wen0, buf\_wen1;} \\ \texttt{\textbf{wire} [7:0] buf\_wv0, buf\_wv1;} \\ \texttt{...} }  \\ \cline{2-4}
			& \makecell[l]{\texttt{\textbf{EPHEMERAL\_STATE}(\textbf{bool}, busy, false)} \\ \texttt{// type, name, reset value} } & \makecell[l]{ \texttt{// reset after each cycle}\\ \texttt{\textbf{bool} busy = false;} } & \texttt{\textbf{wire} busy;}  \\ \hline
			\multirow{2}{*}{\makecell[c]{Transaction \\ Definition \\ (Sec. \ref{sec-method-trans}) } } & \makecell[l]{\texttt{\textbf{REQUEST\_PORT}(query, \textbackslash} \\  \texttt{\ \ \ \ \textbf{ARG}(\textbf{uint8}) idx, \textbackslash} \\ \texttt{\ \ \ \ \textbf{RESP}(\textbf{uint32}) d)} } & \makecell[l]{ \texttt{\textbf{private:} // locally invoked} \\ \texttt{\textbf{bool} query(} \\ \texttt{\ \ \textbf{const uint8\&} idx,} \\ \texttt{\ \ \textbf{uint32\&} d);} } & \makecell[l]{ \texttt{\textbf{output wire} query\_valid,} \\ \texttt{\textbf{output wire} [7:0] query\_idx,} \\ \texttt{\textbf{input wire} [31:0] query\_d,}  \\ \texttt{\textbf{input wire} query\_ready} }  \\ \cline{2-4}
			& \makecell[l]{\texttt{\textbf{SERVICE\_PORT}(query, \textbackslash} \\  \texttt{\ \ \ \ \textbf{ARG}(\textbf{uint8}) idx, \textbackslash} \\ \texttt{\ \ \ \ \textbf{RESP}(\textbf{uint32}) d)} } & \makecell[l]{ \texttt{\textbf{public:} // externally invoked} \\ \texttt{\textbf{bool} query(} \\ \texttt{\ \ \textbf{const uint8\&} idx,} \\ \texttt{\ \ \textbf{uint32\&} d);} } & \makecell[l]{ \texttt{\textbf{input wire} query\_valid,} \\ \texttt{\textbf{input wire} [7:0] query\_idx,} \\ \texttt{\textbf{output wire} [31:0] query\_d,}  \\ \texttt{\textbf{output wire} query\_ready} }  \\ \hline
			\makecell[c]{Logic Block\\ (Sec. \ref{sec-method-logicblock}) } & \texttt{\textbf{LOGIC\_BLOCK}([\textbf{PRIO}(1)])\{c = a + b;\} } & \texttt{\{c = a + b;\}} & \texttt{\textbf{assign} c = a + b;}  \\ \hline
		\end{tabular}
		}
		\label{tab-tpcs-element}
	\end{table*}

	This section defines the modeling abstraction of TPCS. 
	Table \ref{tab-tpcs-element} summarizes the core elements and their correspondence to RTL (SystemVerilog) and C++ representations. 
	As globally visible declarations, a \emph{constant} is a compile-time value and a \emph{struct} is a packed aggregate of named members with static types.
	The remaining elements fall into three categories. 
	First, \emph{modules}, together with their \emph{instances}, \emph{parameters}, and \emph{states}, define the hierarchical structure of a TPCS design. 
	Second, \emph{request} and \emph{service} transaction ports model inter-module communication. 
	Third, \emph{logic blocks} describe intra-module combinational behavior.
	
\subsubsection{Module-Centric Abstraction}
\label{sec-method-module}

	TPCS adopts a module-centric abstraction, representing hardware as a hierarchy of modules. 
	Each module structurally corresponds to an RTL module or a C++ template class. 
	An instance denotes a concrete instantiation of a module within its parent module. 
	Modules are parameterized, with parameters treated as local constants specified at instantiation. 
	This design ensures structural correspondence with both RTL and C++, enabling direct execution and systematic translation.
	
	State elements capture both stored and intermediate values and are categorized into two types. 
	\emph{Persistent state} represents cross-cycle values (analogous to registers) and is updated via deferred assignment, taking effect in the next cycle; it may have multiple assignment ports with different priority. 
	\emph{Ephemeral state} represents intra-cycle temporary values (analogous to wires), which are cleared each cycle and used for sharing intermediate results across logic blocks.

	For a persistent state $r$, let $r_t$ be its value in cycle $t$, and let $W_t=\{(p,d_p)\}$ denote all writes issued in that cycle with \texttt{r.set<p>(d\_p)}, where $p$ denotes the port priority, $d_p$ denotes the data.
	Reads always observe $r_t$.
	If $W_t\neq\emptyset$, the state transition is $r_{t+1}=d_{p^\ast}$, where $p^\ast=\max\{p\mid(p,d_p)\in W_t\}$.
	Otherwise, $r_{t+1}=r_t$.
	Hence, writes are deferred to the next cycle and conflicts are resolved deterministically by priority, corresponding to synchronous RTL non-blocking assignment with priority arbitration.

\subsubsection{Transaction System}
\label{sec-method-trans}


	The transaction system is the central abstraction in TPCS. 
	A \emph{transaction} represents a single-cycle control and data interaction between modules. 
	Transactions may be \emph{handshaked}, where invocation takes effect only when the callee is ready, whereas a non-handshaked transaction is always accepted once invoked.
	
	Modules expose \emph{transaction ports} that define invocable interactions. 
	A \emph{request} port initiates a transaction, while a \emph{service} port responds. 
	Each port optionally specifies exchanged data, including \emph{arguments} from the request side and \emph{responses} from the service side. 
	An invocation implicitly encodes a \emph{valid} signal, while the return value (for handshaked transactions) represents \emph{ready}. 
	
	Ports can be connected when their return types, \emph{arguments}, and \emph{responses} match. 
	Connections are directional: invoking the source port forwards this call to its destination, decoupling port declaration from execution while preserving RTL-style binding semantics. 
	
	Connections must have neither multiple drivers nor dangling ports. 
	Specifically, a child’s \emph{request} port must be connected to a child’s \emph{service}, a parent’s \emph{request}, or bound to logic block(s); 
	a parent’s \emph{service} port must connect to a child’s \emph{service} or be bound to logic block(s). 
	Broadcast is allowed only for transactions without handshake or \emph{responses}; otherwise, connections are one-to-one.
	
\subsubsection{Logic Blocks}
\label{sec-method-logicblock}
	
	Combinational logic in TPCS is expressed as \emph{logic blocks}, implemented as C++ functions. 
	A logic block defines intra-module behavior, including state access, control flow, and transaction invocation. 
	Logic blocks execute only when bound to triggering points, including \emph{service}, triggered by transaction invocation, and \emph{tick}, invoked once per cycle.
	This programmatic form is precise, deterministic, and executable, enabling pre-RTL validation at both module-level and system-level. 
	
	TPCS imposes three constraints to define unambiguous model semantics for subsequent software execution and hardware generation.
	First, each block must terminate and produce deterministic results for all valid inputs, with fully defined control paths, statically bounded loops, statically allocated storage, and no undefined operations.
	Second, conflicting state updates require explicit resolution rules: order-independent blocks use priority-based deferred assignments, whereas explicitly ordered blocks allow later updates to override earlier ones (as shown in Fig. \ref{fig-overview1}).
	Finally, each transaction may be invoked at most once per cycle, where a state assignment port is also treated as a special transaction.

	The single-invocation constraint is enforced at three stages.
	During static validation, all logic blocks within a module are parsed and examined to rule out any feasible execution path that invokes the same transaction more than once.
	During debug simulation, each transaction is guarded by an assert with a clear-on-cycle execution flag.
	During RTL generation, all output signals including transaction signals are checked for multiple drivers.
	
	These constraints define the hardware scope targeted by TPCS.
	Specifically, TPCS models a statically structured subset of deterministic cycle-synchronous hardware, where cycle boundaries and state transitions are explicit and intra-cycle behavior can be represented as terminating combinational computation.
	Asynchronous logic, analog behavior, X/Z states, and unrestricted cross-clock combinational interactions are outside this scope.

\subsubsection{Debugging and Extensions}
	
	\begin{figure} [t]
		\centering
		\includegraphics[]{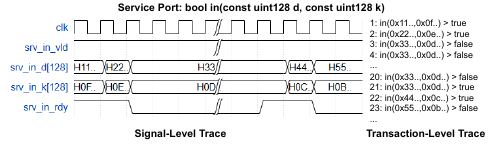}
		\caption{Waveform consistency cross abstraction levels and transaction-to-signal correspondence.}
		\label{fig-wf}
	\end{figure}
	
	TPCS provides built-in tracing for both transactions and registers. 
	As shown in Fig. \ref{fig-wf}, TPCS records traced transaction invocation, input arguments, response values, and return status.
	Traced register updates are captured after deferred assignments are committed. 
	This establishes correspondence between transaction-level trace and signal-level waveform.

	\begin{figure} [t]
		\centering
		\includegraphics[]{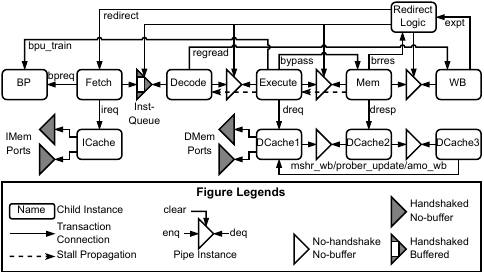}
		\caption{TPCS representation of a simple 5-stage pipelined CPU design using built-in components.}
		\label{fig-pipe5}
	\end{figure}

	TPCS also offers pre-built components for common hardware patterns.
	\emph{Pipe} is a configurable queue instance with \emph{enqueue}, \emph{dequeue}, and \emph{clear} handshaked service ports, and is suitable for pipeline and interconnect modeling.
	Memory components such as \emph{BRAM} support multi-port storage.
	Stall propagation provides a back-pressure mechanism where \emph{stall} transactions are propagated recursively to all connected instances.
	It generates a special transaction port pair together with helper functions that allow logic blocks to access the stall flag and invoke \emph{stall}.
	Fig. \ref{fig-pipe5} shows a 5-stage CPU example where stages are modularized and connected via \emph{Pipe}, with \emph{BRAM} in the cache and stall propagation across the Decode, Execute, and Memory stages.
	
	TPCS supports statically-sized arrayed instances for scalable structured modeling. 
	Point-to-point transaction connections for them implicitly carry index parameters, and bulk connections among adjacent instances can be concisely specified by dimension.
	
\subsection{TPCS Native Execution Semantics}
\label{sec-method-exec}

	This section describes the execution of TPCS models as native C++ programs.
	This section describes transaction invocation, execute-commit simulation, and deterministic parallel execution.
	
	Static execution ordering that allows native control-flow execution without runtime dependency resolution.
	Transaction connections are realized via C++ function calls, eliminating scheduling and activity detection. 
	The evaluation order of logic blocks is fixed at construction time, removing runtime dependency resolution. 
	Due to avoiding global structures such as event queues, TPCS eliminates unnecessary ordering constraints and enables efficient cycle-level parallel simulation.
	
\subsubsection{Transaction Calls and Connections}

	\begin{figure} [t]
		\centering
		\includegraphics[]{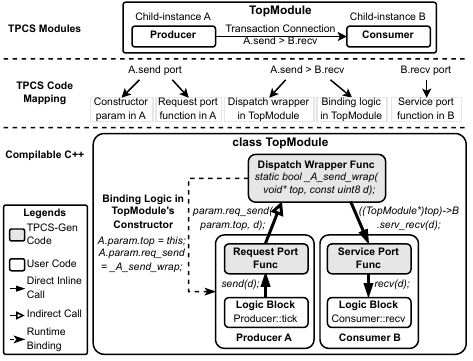}
		\caption{Transforming transaction connection into runtime-bound C++ function pointer call.}
		\label{fig-callpath}
	\end{figure}

	Transaction connections underpin TPCS execution semantics. 
	As shown in Fig. \ref{fig-callpath}, transactions are implemented as C++ function calls via function-pointer-based binding. 
	Both \emph{request} and \emph{service} ports are member functions. 
	The \emph{request} invoked by logic blocks forwards this call through a function pointer, while \emph{service} acts as the entry point.
	Connection is realized by a generated static wrapper function whose pointer is bound to the request side. 
	The wrapper function dispatches this call to connected child instance service or bound logic blocks.
	These codes are automatically derived from the structural specification, requiring only port and connection declarations. 
	
	Without devirtualization, each connection incurs an indirect call that limits inlining.
	When a module type is instantiated under different parents, a shared call site may target different wrappers; optional wrapper duplication trades code size for execution efficiency.
	
	For arrayed child instances, frequent wrapper invocation increases overhead. 
	TPCS mitigates this with an optional aggregation optimization that inlines the child module into parent module, enabling direct calls across siblings. 
	This preserves functional equivalence while trading code size for speedup, and is particularly effective for regular arrays of small processing elements.
	
\subsubsection{Dual-Stage Cycle-Driven Simulation}
	

	Each simulation cycle consists of an \emph{execute} stage and a \emph{commit} stage.
	During the \emph{execute} stage, the \emph{tick} port of each module is invoked, and logic blocks recursively trigger additional request ports as needed to evaluate dependent combinational logic.
	The resulting call structure follows the host's native control flow and directly captures hardware control semantics.
	In the \emph{commit} stage, all deferred assignments are applied and prepared for the next cycle.
	
	Execution naturally follows an ordered traversal of the module hierarchy.
	Each module provides an \emph{exec} entry function, which invokes its own \emph{tick} and all children's \emph{exec} in the order constructed from logic block priorities.
	The \emph{exec} function of the top instance is the entry point of the \emph{execute} stage.
	In contrast, the \emph{commit} stage is order-independent, as state elements are updated individually.

\begin{lstlisting}[
	style=pseudocode,
	caption={Hierarchical Execution Ordering},
	label=alg-seq,
	float=t
	]
Input: Hierarchy H, dependencies {D[i]}, connections
Output: Local execution orders {S[i]}, or FAIL

func LCA(u, v):
  return argmax H.Depth(i) | u, v in H.Subtree(i)

func Project(p, u):
  if p = u return Tick(p)
  return Exec(c) | c in H.Children(p) and u in H.Subtree(c)

owner <- Map reachable block to its initiating Tick's instance
Fail on multiple Tick owners
for each i in H do
 G[i] <- Edgeless graph on Tick(i) and immediate-child Execs
for each i in H do
 for each (a, b) in D[i] do
  u, v <- owner[a], owner[b]; fail if u = v
  p <- LCA(u, v)
  Add edge {Project(p, u) > Project(p, v)} to G[p]
return {TopologicalSort(G[i])}
\end{lstlisting}

	The execution-order construction based on call ownership and hierarchical projection is shown in Algorithm \ref{alg-seq}.
	Variables follow the convention that $i$ denotes an instance, $p$ a parent or ancestor, and $c$ an immediate child.
	Here, $H$ denotes the instance hierarchy, $D[i]$ contains the precedence pairs of logic blocks within $i$, and $S[i]$ is the resulting local execution sequence in $i$.
	The algorithm assumes resolved, nonrecursive calls without repeated block invocation, and dependency pairs whose endpoints both have owners.
	Unresolved calls, conflicting ownership, or cyclic ordering constraints cause a design fatal.
	Logic blocks without specified priority impose no additional ordering.
	Algorithm \ref{alg-seq} takes $O(B+C+N+Dh)$ time with visited-node tracking and parent-pointer traversal, where $B$, $C$, $N$, $D$, and $h$ denote the numbers of logic blocks, call edges, module instances, dependency pairs, and the hierarchy depth, respectively.
	
	Correctness rests on projecting each block dependency to the lowest common ancestor of its owning instances.
	Because child Execs execute indivisibly, recursive composition of the local topological orders preserves all declared dependencies.
	Dependencies within a single Tick are conservatively rejected, as external scheduling cannot alter its internal order.
	This ordering ensures that the actual execution respects the declared logic-block precedence in every module instance, thereby making software execution consistent with the model's declared semantics.

\subsubsection{Parallel Execution}
	
	\begin{figure} [t]
		\centering
		\includegraphics[]{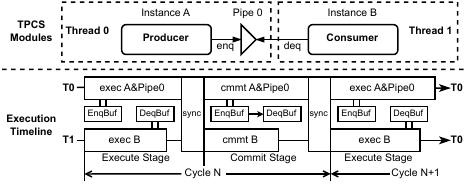}
		\caption{TPCS multi-thread parallel execution and synchronization.}
		\label{fig-multithread}
	\end{figure}

	

	TPCS adopts static parallelization at component boundaries (Fig. \ref{fig-multithread}).
	Only order-free instances may be assigned to different worker threads, and the partition is fixed before simulation and remains unchanged during execution.
	All persistent and ephemeral states of an instance are exclusively owned and written by its assigned thread.
	Cross-thread interactions are restricted to \emph{pipe} or side-effect-free services that only read persistent states; ordinary service/request interactions with cross-thread side effects are disallowed.

	Cross-thread \emph{pipe} communication follows the execute--commit semantics.
	During \emph{execute}, the producer and consumer independently update thread-local enqueue and dequeue buffers, while persistent pipe state remains unchanged.
	After an acquire/release spin barrier, the enqueue-side owner deterministically commits both endpoints to the canonical pipe state.
	A second barrier separates commit from the next execute stage, so transferred data becomes visible only in the following cycle.
	Therefore, worker interleaving cannot introduce data races or affect cycle-boundary state: each thread writes only private state during execution, while all cross-thread effects are applied at deterministic synchronization points.
	
	The per-thread execution order and instance-to-thread mapping are fully static, with no work stealing, randomized scheduling, or runtime arbitration; hence no simulator-side random mechanism is required for determinism.
	Thread pinning is used only to reduce performance variance and is not required for correctness.
	The half-cycle barriers introduce synchronization overhead, so parallel execution is most effective for sufficiently large designs and fewer cross-thread pipes.

\subsection{Synthesizable RTL Generation from TPCS}
\label{sec-method-rtl}

	The goal of RTL generation in TPCS is to translate a high-level model into synthesizable RTL while preserving structural correspondence and execution semantics.
	Accordingly, TPCS adopts a generation strategy that decouples structural components from logic code.
	Module hierarchy, state elements, and transaction ports can be directly translated to form the RTL skeleton. 
	In contrast, logic blocks and transaction invocations are transformed into signal-level combinational logic via a dedicated LLVM-based compiler.
	
\subsubsection{RTL Generation Workflow}

	\begin{figure} [t]
		\centering
		\includegraphics[]{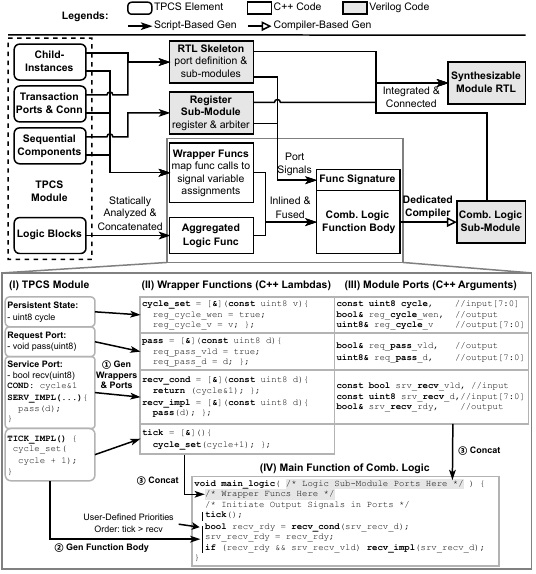}
		\caption{RTL Generation Flow for the TPCS Model.}
		\label{fig-gen}
	\end{figure}
	
	RTL generation in TPCS is performed per module following the model hierarchy. 
	As shown in Fig. \ref{fig-gen}, each module is decomposed into three components: 
	(1) a \emph{Register} sub-module encapsulating persistent state with write arbitration, 
	(2) a \emph{Logic} sub-module implementing combinational behavior, and 
	(3) an RTL skeleton with module port definitions, child-instances, and interconnections.
	Only the \emph{Logic} sub-module is generated via dedicated compiler; all other structures are derived directly from the TPCS specification. 
	This decomposition separates sequential and combinational logic, aligning with TPCS dual-stage execution semantics.
	
	For \emph{Logic} sub-module, to unify interactions as port-level interfaces, a void-return function is constructed where all ports are explicit arguments. 
	(1) TPCS elements are transformed into wrapper functions and port definitions, where wrappers preserve APIs but translate calls into port accesses (e.g., \emph{xxx\_set} sets register write signals).
	(2) Logic blocks are invoked sequentially in user-defined priorities. \emph{Tick}-bound blocks always execute, while \emph{service}-bound blocks are conditionally triggered by \emph{valid} and \emph{ready} signals.
	(3) All components are composed into a single function, whose execution aggregates all port-level reads and writes within one cycle.
	
\subsubsection{Dedicated Compiler for Combinational Logic Generation}
	
	\begin{figure} [t]
		\centering
		\includegraphics[]{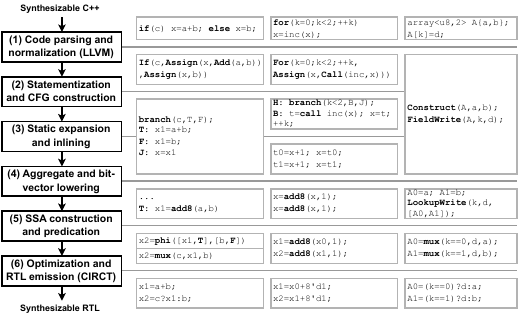}
		\caption{Representative transformations through the six stages in TPCS compiler.}
		\label{fig-genbackend}
	\end{figure}

	The TPCS compiler translates combinational C++ functions into synthesizable SystemVerilog.
	It supports Boolean and fixed-width integer operations, static arrays and aggregates, conditional control flow, fully unrollable loops, and statically resolved, nonrecursive calls.
	Dynamic allocation, runtime dispatch, native floating-point operations, and unsupported library calls are excluded.
	State and cycle boundaries remain explicit in TPCS.
	
	Fig. \ref{fig-genbackend} illustrates six stages: semantic normalization using LLVM, statementization and Control Flow Graph (CFG) construction, expansion and inlining, aggregate and bit-vector lowering, Static Single Assignment (SSA) construction and predication, and optimized RTL emission using CIRCT \footnote{https://github.com/llvm/circt}.
	These stages expose data dependencies, preserve evaluation order and width semantics, and eliminate loops and calls without introducing additional state.
	
	Predication converts control flow into value selection.
	Let $g_B$ denote block reachability and $e_{B,J}$ the predicate for traversing edge $B\rightarrow J$.
	For a branch on $c$,
	\begin{equation}
		\begin{aligned}
			g_{\mathrm{entry}} &= 1, &
			g_J &= \bigvee_{B\in\operatorname{Pred}(J)} e_{B,J}, \\
			e_{B,T} &= g_B\land c, &
			e_{B,F} &= g_B\land\neg c .
		\end{aligned}
		\label{eq:rtl-predicates}
	\end{equation}
	An SSA merge pairs each value with its predecessor and lowers to
	\begin{equation}
		\phi([v_1,B_1],[v_2,B_2])
		\;\leadsto\;
		\operatorname{mux}(e_{B_1,J},v_1,v_2),
		\label{eq:rtl-phi}
	\end{equation}
	where $\leadsto$ denotes lowering and $\operatorname{mux}(s,u,v)$ selects $u$ if $s$ holds and $v$ otherwise.
	Incoming predicates are mutually exclusive and cover execution of $J$.
	A guarded write $\texttt{if}(g)\ A[k]:=d$ similarly becomes
	\begin{equation}
		A_j^{\mathrm{new}}=\operatorname{mux}(g\land(k=j),d,A_j^{\mathrm{old}}),
		\quad 0\le j<N.
		\label{eq:rtl-array-write}
	\end{equation}
	SSA versions preserve sequential updates as combinational dependencies, while predicate checks enforce valid uses and complete output coverage.
	These rules preserve values along the acyclic dependency graph, making settled RTL outputs equivalent to one C++ execution for valid inputs.

\subsubsection{Predicate Sinking Optimization}

\begin{lstlisting}[
	style=pseudocode,
	caption={Demand-Guided Predicate Sinking},
	label=alg-pred,
	float=t
	]
Input: SSA signal-driving DAG G
Output: Optimized DAG G
	
func FirstGatingPredicates(x):
  P <- {}; Q <- ConsumerEdges(x)
  while Q is not empty do
    // visited check are omitted here
    e <- Pop(Q); u <- Consumer(e)
    if u is output node return FAIL
    if u is Mux and u gates e then
      P <- P union {DemandPredicate(u, e)}
    else
      Q <- Q union ConsumerEdges(u)
  return P

for each x = Mux(g, t, f) in RevTopoOrder(G) do
  P <- FirstGatingPredicates(x)
  if P = FAIL continue
  for each q in BFS(PredicateCone(g)) do
    if forall p in P: p => q then
      g' <- (Clone(g) where [q <- True])
      x <- Mux(g', t, f); break
    if forall p in P: p => not q then
      g' <- (Clone(g) where [q <- False])
      x <- Mux(g', t, f); break
return G
\end{lstlisting}

	A key challenge in lowering TPCS C++ combinational logic into synthesizable RTL is the serial predicate chain introduced by nested control flow.
	After predication, intermediate values may be repeatedly guarded by accumulated branch conditions, increasing control dependence and combinational depth.
	In handwritten RTL, similar logic is often expressed in an \textbf{always\_comb} block, where data computation and predicate evaluation proceed largely in parallel and predicates are applied only near observable boundaries.
	To recover this structure, the TPCS compiler introduces a demand-guided predicate-sinking optimization before entering the CIRCT back-end.
	
	The procedure is shown in Algorithm \ref{alg-pred}.
	For a candidate $x=\operatorname{Mux}(g,t,f)$, \textsc{FirstGatingPredicates} follows each consumer path to the first Mux that conditionally requires $x$ and collects its demand predicate into $P$.
	A true-arm use contributes $g'$, while a false-arm use contributes $\neg g'$.
	If any path reaches an observable boundary without such gating, the candidate is rejected.
	Condition uses and uses through both data arms are likewise treated as ungated.
	
	The predicate cone of $g$ is then searched for a node $q$ such that
	$
	\forall p\in P; p\Rightarrow q
	$
	or
	$
	\forall p\in P; p\Rightarrow \neg q.
	$
	In these cases, $q$ is constant whenever $x$ is demanded and can be replaced by \texttt{True} or \texttt{False} in a cloned guard cone, followed by Boolean simplification.
	The original cone is preserved for shared uses.
	
	With visited-edge tracking, one pass takes $O(V+E+M[(V+E)+KRC_{\mathrm{proof}}+C_{\mathrm{clone}}])$ time, where $V,E$ bound the DAG size during the pass, $M$ counts candidate muxes, $K,R$ bound predicate-cone and demand-set sizes, and $C_{\mathrm{clone}}$ bounds cloning and simplification.
	An exhaustive implication check on $s$ nodes and $k$ Boolean inputs costs $C_{\mathrm{proof}}=O(s2^k)$; practical proof budgets bound search effort, with unproven candidates left unchanged.
	
	This is especially useful for temporaries inside nested \texttt{if} statements.
	For example, if a temporary is guarded by $a\land b$, but every downstream use already occurs under a condition implying $a$, then $a$ is redundant at that temporary.
	Predicate sinking specializes the guard from $a\land b$ to $b$, removing branch conditions already enforced downstream and shortening the resulting predicate chain without changing observable output.
	
	Before predicate sinking, Boolean signal networks are normalized into And-Inverter Graphs (AIGs) and simplified using structural rules and truth-table-based reduction, which reduces the cost of predicate proofs.
	Other common optimizations are performed by \texttt{circt-opt}, with the same configuration as Chisel release-mode, including canonicalization, common-subexpression elimination, dead-code elimination, and hardware-level cleanup passes.

\section{Experiment and Validation}

	We evaluate TPCS through four complementary experiments:
	\begin{itemize}
		\item \textbf{Cross-abstraction correctness (Section \ref{sec-exp-correct})}: Compare waveforms and workload results between direct execution and generated RTL to verify that simulation gains preserve the evaluated cycle behavior.
		\item \textbf{System-level performance (Section \ref{sec-exp-sys})}: Measure SMP simulation speed, parallel scalability, and overheads to assess efficiency on large-scale practical systems.
		\item \textbf{Direct HDL comparison (Section \ref{sec-exp-cmpr})}: Hold microarchitecture constant to compare simulation speed and RTL quality across existing HDL workflows, and use ablation to assess dedicated compiler optimization.
		\item \textbf{Hardware applicability (Section \ref{sec-exp-asic})}: Evaluate AES and systolic-array designs under different input activity and transaction granularity to test effectiveness beyond CPU pipelines.
	\end{itemize}
	
\subsection{Experiment Setup}

\subsubsection{Platforms and Toolchain}
	
	All experiments are conducted on a PC equipped with a Intel i9-14900K processor and 128 GB of DDR5 memory, running Ubuntu 22.04.
	Verilator 5.036 is used as the baseline in all experiments.
	Both TPCS code and Verilator-based simulation code are compiled using g++ 11.4 with the C++20 standard and -O3 optimization.
	The TPCS RTL generation scripts are executed on Python 3.11, with LLVM/clang 18.0 as TPCS compiler front-end.

	\begin{figure} [t]
		\centering
		\subfigure[RISC-V SMP system.]{
			\begin{minipage}{\columnwidth}
				\centering
				\includegraphics[]{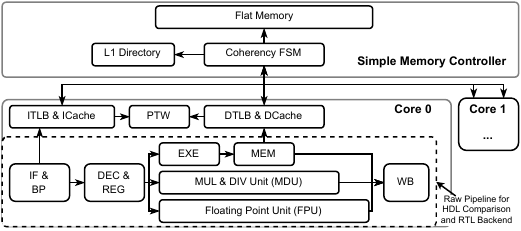}
			\end{minipage}
			\label{fig-hw-cpu}
		}
		\subfigure[AES Enc unit.]{
			\begin{minipage}{0.27\columnwidth}
				\centering
				\includegraphics[]{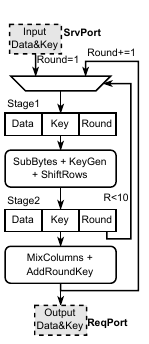}
			\end{minipage}
			\label{fig-hw-aes}
		}
		\subfigure[Systolic array.]{
			\begin{minipage}{0.65\columnwidth}
				\centering
				\includegraphics[]{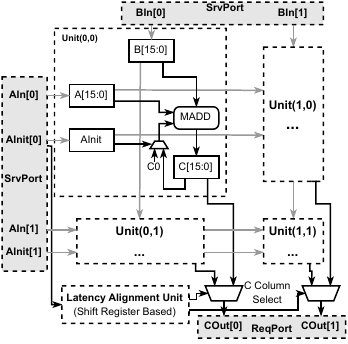}
			\end{minipage}
			\label{fig-hw-array}
		}
		\caption{Representative target $\mu$arch designs used in evaluation.}
		\label{fig-hw}
	\end{figure}
	
\subsubsection{Target Hardware Design}

	Fig. \ref{fig-hw} illustrates the $\mu$arch details of the three target hardware designs.
	The scalable RISC-V SMP system adopts a standard 5-stage in-order pipeline, similar to Rocket \cite{ref-chipyard}, with BTB, BHT, RAS predictors and a 32KB 4-way L1 cache.
	A lightweight atomic cache system is used as the memory subsystem.
	The ASIC targets, including an AES encryption unit and a systolic array, follow conventional dataflow architectures.
	The AES unit is implemented with one or two pipeline stages (denoted as \textbf{AES1} and \textbf{AES2}), while the systolic array is instantiated in two sizes, 8x8 and 16x16 (denoted as \textbf{SA8} and \textbf{SA16}).
	
	The RISC-V SMP system executes workloads pre-loaded into memory, while the accelerator designs are driven by a test harness providing streaming inputs and collecting outputs.
	The harness is shared between TPCS and Verilator, differing only at the transaction or signal interface.
	
	For the SMP and accelerator simulation experiments, TPCS execution is compared with Verilator simulation of RTL generated from the same model.
	Handcrafted SystemVerilog implementations provide references for RTL quality evaluation, preserving the same module hierarchy and register definitions.
	Section \ref{sec-exp-cmpr} additionally compares handcrafted SV and Chisel, SystemC, PyMTL, and TPCS implementations of the same CPU pipeline in both simulation performance and RTL quality.
	
\subsubsection{Workloads}
\label{sec-exp-setup-workload}

	For the RISC-V SMP system, \textbf{CoreMark} and \textbf{RISC-V bmarks} bare-metal benchmarks are used.
	A custom checkpoint instruction signals workload start and end, implemented as a dedicated pull-up signal (or transaction) connected to the top-level module.
	This mechanism is used only for timing measurement.
	
	For the accelerator targets, two input patterns are evaluated.
	In the \textbf{Full} mode, data is continuously supplied to fully utilize the pipeline.
	In the \textbf{Burst} mode, input is periodically stalled to emulate intermittent availability: after every 10 input batches, the stream pauses for an equal duration.
	For example, pauses last 100 cycles for AES (10 cycles per batch) and 80 cycles for an 8x8 SA (8 cycles per matrix).
	Identical input patterns are applied across all implementations.

\subsubsection{Simulation Configuration}

	For correctness validation, waveform-level comparisons are performed.
	In the RISC-V core, valid signals and PCs of each pipeline stage are traced, while for accelerators, top-level I/O ports are monitored.
	
	Simulation speed is measured in simulated cycles per second, with waveform dumping disabled during performance measurements.
	The SMP experiments measure benchmark execution phases using checkpoint timing; the direct pipeline comparison follows the fixed-workload protocol in Section \ref{sec-exp-cmpr}, and accelerator measurements use continuous 10-minute execution windows.
	Each comparison uses identical workloads, inputs, and initial states across implementations.
	
	For multithreaded RISC-V SMP simulation, Verilator uses its default parallel strategy, while TPCS evenly distributes cores across threads.
	The simplified memory controller is assigned to the first TPCS worker thread.
	All threads are pinned to physical cores to reduce scheduling overhead.

\subsubsection{RTL Backend Flow}

	RTL quality is evaluated after routing using OpenROAD and ASAP7, reporting minimum clock period and standard-cell area excluding filler cells.
	Clock constraints are selected through timing sweeps and held identical across implementations of each target, together with the remaining backend settings.
	Multiple clock constraints are explored, and the minimum achievable clock period after place-and-route is reported.
	
	For the RISC-V SMP system, only the isolated pipeline is synthesized to avoid bias from large memory structures.
	This ensures the evaluation focuses on logic quality rather than memory-dominated effects.
	
\subsection{Validation on Cross-Abstraction-Level Correctness}
\label{sec-exp-correct}
	
	\begin{table}[t]
		\centering
		\caption{Cycle-Level Equivalence Between TPCS Execution and Generated RTL Simulation.}
		\resizebox{0.95\linewidth}{!}{
		\begin{tabular}{|c|c|c|c|}
			\hline
			\textbf{\begin{tabular}[c]{@{}c@{}}Design \& Workload\end{tabular}}            & \textbf{\begin{tabular}[c]{@{}c@{}}Observed Registers \\ \&  Signals\end{tabular}}       & \textbf{\begin{tabular}[c]{@{}c@{}}Simulated\\Cycles\end{tabular}} & \textbf{\begin{tabular}[c]{@{}c@{}}Waveform\\ Equivalence\end{tabular}} \\ \hline
			\begin{tabular}[c]{@{}c@{}}RISC-V 1-Core\\ CoreMark 1-Iter\end{tabular}          & \begin{tabular}[c]{@{}c@{}}PCReg of each stage\\ \& IReq \& DReq\end{tabular} & 453.6K                & Exact                \\ \hline
			\begin{tabular}[c]{@{}c@{}}AES Enc (2-Piped)\\ Full Mode 1K Packs\end{tabular}   & \begin{tabular}[c]{@{}c@{}}InData \& Inkey \& InRdy\\ \& OutData\end{tabular}  & 10.0K                 & Exact                  \\ \hline
			\begin{tabular}[c]{@{}c@{}}AES Enc (2-Piped)\\ Burst Mode 1K Packs\end{tabular}  & \begin{tabular}[c]{@{}c@{}}InData \& Inkey \& InRdy\\ \& OutData\end{tabular}  & 18.2K                 & Exact                  \\ \hline
			\begin{tabular}[c]{@{}c@{}}Systolic Array 8x8\\ Full Mode 1K Matrices\end{tabular}  & \begin{tabular}[c]{@{}c@{}}AIn \& AInit \& BIn\\ \& InRdy \& COut\end{tabular}  & 8.0K                  & Exact                  \\ \hline
			\begin{tabular}[c]{@{}c@{}}Systolic Array 8x8\\ Burst Mode 1K Matrices\end{tabular} & \begin{tabular}[c]{@{}c@{}}AIn \& AInit \& BIn\\ \& InRdy \& COut\end{tabular}  & 15.3K                 & Exact                  \\ \hline
		\end{tabular}
		}
		\label{tab-sig}
	\end{table}
	
	\begin{figure} [t]
		\centering
		\includegraphics[]{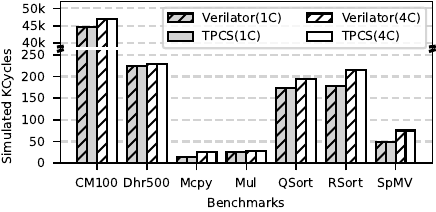}
		\caption{Simulated cycle numbers for running each workload on target RISC-V systems. 4C denotes 4-core SMP systems running 4 copies of the workload.}
		\label{fig-acc}
	\end{figure}
	
	Table \ref{tab-sig} compares module-interface signals and selected registers cycle by cycle between direct TPCS execution and Verilator simulation of its generated RTL.
	All observed values match throughout the traced CPU, AES, and systolic-array executions.
	RTL tracing begins after reset, with corresponding TPCS states initialized through C++ object construction.

	Fig. \ref{fig-acc} further shows identical workload cycle counts for TPCS and RTL simulation on single-core and four-core SMP systems.
	Both reproduce the limited scaling of memory-intensive workloads such as memcpy and spmv under contention in the simplified memory system.
	These observations validate agreement at both signal and workload levels for the evaluated configurations.

	All subsequent comparative experiments require matching cycle counts and identical architecturally visible states across implementations as baseline acceptance criteria. Cycle-by-cycle register-waveform agreement is checked directly in experiments without performance measurements and in additional untimed runs with waveform tracing enabled for performance experiments.

\subsection{Validation on System-Level Simulation Performance}
\label{sec-exp-sys}
	
	\begin{figure} [t]
		\centering
		\subfigure[1-core. $40\times$ speedup.]{
			\begin{minipage}{\columnwidth}
				\centering
				\includegraphics[]{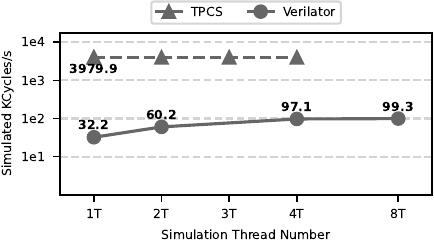}
			\end{minipage}
			\label{fig-effi-cpu1}
		}
		\subfigure[4-cores. $42\times$ speedup.]{
			\begin{minipage}{\columnwidth}
				\centering
				\includegraphics[]{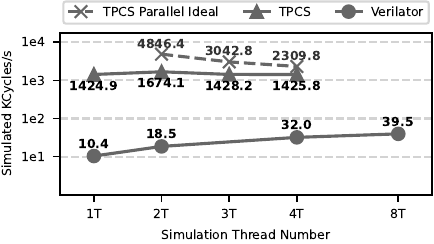}
			\end{minipage}
			\label{fig-effi-cpu2}
		}
		\subfigure[8-cores. $36\times$ speedup.]{
			\begin{minipage}{\columnwidth}
				\centering
				\includegraphics[]{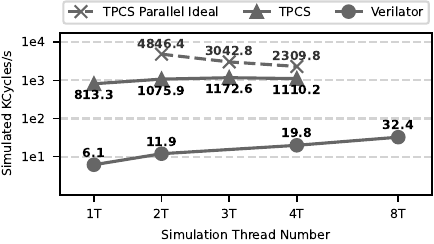}
			\end{minipage}
			\label{fig-effi-cpu3}
		}
		\caption{Simulation speed comparison on different target $\mu$arch configurations with N-copies CoreMark workloads.}
		\label{fig-effi-cpu}
	\end{figure}

	Fig. \ref{fig-effi-cpu} compares the simulation speed of TPCS and Verilator on RISC-V SMP systems with varying numbers of cores.
	The consistency of output results, cycle counts, and architectural states are checked for each run across simulation thread numbers.
	Simulation speed is measured as the number of simulated cycles per second when running 100 iterations of CoreMark, and speedup is reported based on the maximum performance achieved under different simulation thread numbers.
	Results with 1-8 simulation threads are shown, as both TPCS and Verilator exhibit limited scalability beyond this range.
	The dashed line indicates the theoretical upper bound of TPCS parallel simulation speed, constrained by synchronization overhead.
	
	Overall, TPCS achieves a 36$\times$-42$\times$ speedup over the RTL simulation baseline on this control-flow-dominated CPU target hardware.
	For 4-core and 8-core configurations, peak performance of parallel TPCS is reached at 2 and 3 threads, respectively, indicating that faster per-thread execution increases the relative cost of synchronization.
	
	The speedup mainly arises from the significant amount of inactive logic in CPU pipelines.
	Pipeline stalls may be caused by memory accesses or branch mispredictions, and each instruction activates only a subset of functional units in each stage, which creates substantial opportunities for pruning in TPCS.
	In contrast, dataflow-driven ASIC designs exhibit more regular execution with higher utilization, resulting in smaller speedups.
	
	\begin{figure} [t]
		\centering
		\includegraphics[]{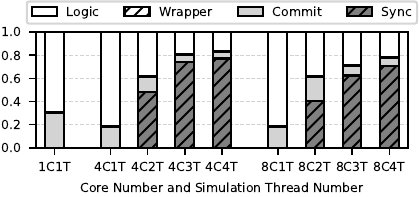}
		\caption{Overhead composition of TPCS execution on RISC-V SMP target.}
		\label{fig-compos1}
	\end{figure}
	
	Fig. \ref{fig-compos1} further presents a breakdown of TPCS simulation overhead on the RISC-V SMP system. Measurements are collected using the \emph{perf} tool with call tracing enabled, excluding initialization and teardown phases.
	The execution phase is further divided into logic evaluation and transaction wrapper overhead, where "Logic" accounts for the execution of logic blocks and "Wrapper" captures the cost of transaction interface handling.
	"Commit" represents the commit stage, and "Sync" represents the synchronization at each half-cycle.
	
	As the thread count increases, the ratio between the commit and execution stages rises because inter-thread pipes reduce data locality and introduce additional cache transfer overhead in the commit stage.
	At the same time, the proportion of synchronization time grows significantly because the per-cycle computation time after task partitioning does not substantially exceed the synchronization cost.

	\begin{table}[t]
		\centering
		\caption{Qualitative comparison with prior simulation frameworks.}
			\begin{tabular}{|c|cccc|}
				\hline
				Framework                                                                              & \multicolumn{1}{c|}{\textbf{GSIM}}  & \multicolumn{1}{c|}{\textbf{TPCS}}  & \multicolumn{1}{c|}{\textbf{Assassyn}} & \textbf{GEM5} \\ \hline
				\textbf{Abstraction Level}                                                             & \multicolumn{1}{c|}{RTL sim}        & \multicolumn{2}{c|}{SW-modeled HW}                                           & Arch sim      \\ \hline
				\textbf{Target CPU Design}                                                             & \multicolumn{2}{c|}{RV64 5-pipelined}                                   & \multicolumn{2}{c|}{RV32 5-pipelined}                 \\ \hline
				\textbf{\begin{tabular}[c]{@{}c@{}}Speedup to Single-\\ Threaded Verilator\end{tabular}} & \multicolumn{1}{c|}{$\sim$20$\times$}      & \multicolumn{1}{c|}{$\sim$100$\times$}     & \multicolumn{1}{c|}{$\sim$3$\times$}          & $\sim$40$\times$     \\ \hline
				\textbf{\begin{tabular}[c]{@{}c@{}}Speedup to\\ Parallel Verilator\end{tabular}}       & \multicolumn{1}{c|}{$\sim$6$\times$}       & \multicolumn{1}{c|}{$\sim$40$\times$}      & \multicolumn{2}{c|}{Not reported}                       \\ \hline
				\textbf{Cycle Error Rate}                                                              & \multicolumn{1}{c|}{0} & \multicolumn{1}{c|}{0} & \multicolumn{1}{c|}{0}    & 2-20\%        \\ \hline
				\textbf{Data Reported By}                                                                   & \multicolumn{1}{c|}{GSIM \cite{ref-gsim}}           & \multicolumn{1}{c|}{This work}      & \multicolumn{2}{c|}{Assassyn \cite{ref-assassyn}}                          \\ \hline
			\end{tabular}
		\label{tab-ref-comp}
	\end{table}

	Table \ref{tab-ref-comp} provides performance data reported from published GSIM, Assassyn, and GEM5 results \cite{ref-assassyn, ref-gsim} for qualitative comparison.
	Differences in target designs, modeling scope, and baseline configurations preclude direct performance ranking; Section \ref{sec-exp-cmpr} provides a controlled comparison on an identical pipeline cross synthesizable modeling methods.

\subsection{Direct Comparison with Existing Synthesizable HDLs}
\label{sec-exp-cmpr}

{
\setlength{\tabcolsep}{1pt}
\begin{table}[t]
	\centering
	\caption{Direct Comparison with Existing Synthesizable HDLs.}
	\resizebox{\linewidth}{!}{
	\begin{tabular}{|c|cc|ccc|ccc|}
		\hline
		\multirow{2}{*}{\textbf{Language}}                                 & \multicolumn{2}{c|}{\textbf{Sim Speed (MCycles/s)}}                                                                                                          & \multicolumn{3}{c|}{\textbf{Minimum Clock Period (ps)}}                                                                                                                                                                                                     & \multicolumn{3}{c|}{\textbf{Standard-Cell Area (um\textasciicircum{}2)}}                                                                                                                                                                             \\ \cline{2-9} 
		& \multicolumn{1}{c|}{\textbf{\begin{tabular}[c]{@{}c@{}}Direct\\ Sim\end{tabular}}}   & \textbf{\begin{tabular}[c]{@{}c@{}}RTLGen+\\ Verilator\end{tabular}} & \multicolumn{1}{c|}{\textbf{\begin{tabular}[c]{@{}c@{}}Full\\ Pipeline\end{tabular}}}     & \multicolumn{1}{c|}{\textbf{\begin{tabular}[c]{@{}c@{}}MDU-\\ Only\end{tabular}}}        & \textbf{\begin{tabular}[c]{@{}c@{}}FPU-\\ Only\end{tabular}}         & \multicolumn{1}{c|}{\textbf{\begin{tabular}[c]{@{}c@{}}Full\\ Pipeline\end{tabular}}}   & \multicolumn{1}{c|}{\textbf{\begin{tabular}[c]{@{}c@{}}MDU-\\ Only\end{tabular}}}      & \textbf{\begin{tabular}[c]{@{}c@{}}FPU-\\ Only\end{tabular}}      \\ \hline
		\begin{tabular}[c]{@{}c@{}}Handcrafted\\ SV\end{tabular}           & \multicolumn{2}{c|}{\begin{tabular}[c]{@{}c@{}}2.384\\ (ref 1.0)\end{tabular}}                                                                               & \multicolumn{1}{c|}{\begin{tabular}[c]{@{}c@{}}1310.39\\ (ref 1.0)\end{tabular}}          & \multicolumn{1}{c|}{\begin{tabular}[c]{@{}c@{}}1028.30\\ (ref 1.0)\end{tabular}}         & \begin{tabular}[c]{@{}c@{}}1092.35\\ (ref 1.0)\end{tabular}          & \multicolumn{1}{c|}{\begin{tabular}[c]{@{}c@{}}20438\\ (ref 1.0)\end{tabular}}          & \multicolumn{1}{c|}{\begin{tabular}[c]{@{}c@{}}1192\\ (ref 1.0)\end{tabular}}          & \begin{tabular}[c]{@{}c@{}}6498\\ (ref 1.0)\end{tabular}          \\ \hline
		\begin{tabular}[c]{@{}c@{}}Chisel with\\ GSIM\end{tabular}         & \multicolumn{1}{c|}{\begin{tabular}[c]{@{}c@{}}4.218\\ (x1.8)\end{tabular}}           & \begin{tabular}[c]{@{}c@{}}2.027\\ (x0.85)\end{tabular}              & \multicolumn{1}{c|}{\begin{tabular}[c]{@{}c@{}}1296.70\\ (-1.05\%)\end{tabular}}          & \multicolumn{1}{c|}{\begin{tabular}[c]{@{}c@{}}988.23\\ (-3.90\%)\end{tabular}}          & \begin{tabular}[c]{@{}c@{}}1217.23\\ (+11.4\%)\end{tabular}          & \multicolumn{1}{c|}{\begin{tabular}[c]{@{}c@{}}19842\\ (-2.92\%)\end{tabular}}          & \multicolumn{1}{c|}{\begin{tabular}[c]{@{}c@{}}1190\\ (-0.17\%)\end{tabular}}          & \begin{tabular}[c]{@{}c@{}}6513\\ (+0.23\%)\end{tabular}          \\ \hline
		\begin{tabular}[c]{@{}c@{}}Synthesizable\\ SystemC\end{tabular}     & \multicolumn{1}{c|}{\begin{tabular}[c]{@{}c@{}}0.772\\ (x0.32)\end{tabular}}          & \begin{tabular}[c]{@{}c@{}}2.329\\ (x0.98)\end{tabular}              & \multicolumn{1}{c|}{\begin{tabular}[c]{@{}c@{}}1331.24\\ (+1.59\%)\end{tabular}}          & \multicolumn{1}{c|}{\begin{tabular}[c]{@{}c@{}}1061.63\\ (+3.24\%)\end{tabular}}         & \begin{tabular}[c]{@{}c@{}}1307.12\\ (+19.7\%)\end{tabular}          & \multicolumn{1}{c|}{\begin{tabular}[c]{@{}c@{}}22792\\ (+11.5\%)\end{tabular}}          & \multicolumn{1}{c|}{\begin{tabular}[c]{@{}c@{}}1282\\ (+7.55\%)\end{tabular}}          & \begin{tabular}[c]{@{}c@{}}7901\\ (+21.6\%)\end{tabular}          \\ \hline
		PyMTL                                                              & \multicolumn{1}{c|}{\begin{tabular}[c]{@{}c@{}}\textless 0.01\\ (\textless x0.01)\end{tabular}}         & \begin{tabular}[c]{@{}c@{}}2.130\\ (x0.89)\end{tabular}              & \multicolumn{1}{c|}{\begin{tabular}[c]{@{}c@{}}1329.20\\ (+1.44\%)\end{tabular}}          & \multicolumn{1}{c|}{\begin{tabular}[c]{@{}c@{}}1073.39\\ (+4.39\%)\end{tabular}}         & \begin{tabular}[c]{@{}c@{}}1137.83\\ (+4.16\%)\end{tabular}          & \multicolumn{1}{c|}{\begin{tabular}[c]{@{}c@{}}20263\\ (-0.86\%)\end{tabular}}          & \multicolumn{1}{c|}{\begin{tabular}[c]{@{}c@{}}1183\\ (-0.76\%)\end{tabular}}          & \begin{tabular}[c]{@{}c@{}}6386\\ (-1.72\%)\end{tabular}          \\ \hline
		\textbf{\begin{tabular}[c]{@{}c@{}}TPCS\\ Native C++\end{tabular}} & \multicolumn{1}{c|}{\textbf{\begin{tabular}[c]{@{}c@{}}11.37\\ (x4.77)\end{tabular}}} & \textbf{\begin{tabular}[c]{@{}c@{}}2.168\\ (x0.91)\end{tabular}}     & \multicolumn{1}{c|}{\textbf{\begin{tabular}[c]{@{}c@{}}1269.84\\ (-3.09\%)\end{tabular}}} & \multicolumn{1}{c|}{\textbf{\begin{tabular}[c]{@{}c@{}}979.58\\ (-4.74\%)\end{tabular}}} & \textbf{\begin{tabular}[c]{@{}c@{}}1120.50\\ (+2.58\%)\end{tabular}} & \multicolumn{1}{c|}{\textbf{\begin{tabular}[c]{@{}c@{}}19761\\ (-3.31\%)\end{tabular}}} & \multicolumn{1}{c|}{\textbf{\begin{tabular}[c]{@{}c@{}}1212\\ (+1.68\%)\end{tabular}}} & \textbf{\begin{tabular}[c]{@{}c@{}}6743\\ (+3.77\%)\end{tabular}} \\ \hline
	\end{tabular}
	}
	\label{tab-direct-comp}
\end{table}
}

	Table \ref{tab-direct-comp} compares implementations of the same RV64GC pipeline extracted from the SMP system, including branch prediction but excluding the memory subsystem, with privileged and exception handling disabled.
	All implementations preserve identical register definitions and cycle behavior, using idiomatic descriptions and timing-oriented optimizations at the modeling level.
	Chisel 7.15.0 uses CHIRRTL-GSIM \footnote{GSIM commit ed45bc9, September 22, 2026} simulation and CIRCT RTL generation; SystemC uses libsystemc 3.0.1 and the Intel Compiler for SystemC (ICSC); PyMTL 3.1.17 uses pure-Python simulation under Python 3.11.

\subsubsection{On Simulation Speed}

	All implementations execute ten bare-metal CoreMark iterations under the same harness, taking 8,970,967 cycles and retiring 3,681,166 instructions, timed from reset completion to program completion.
	Simulations are single-threaded with C++ compiled using \texttt{-O3}; results average ten runs with less than 1\% variation, except for pure-Python PyMTL, measured once.
	TPCS reaches 11.37 MCycles/s, outperforming handcrafted SV under Verilator, Chisel under GSIM, and native SystemC by $4.77\times$, $2.70\times$, and $14.73\times$, respectively.
	As reference, generated RTL achieves 2.027--2.329 MCycles/s under the same Verilator version.
	
	The isolated pipeline simulates faster but shows a smaller relative speedup than the SMP system.
	Removing caches and multicore interactions reduces simulation work, while single-cycle memory accesses increase pipeline utilization and limit pruning opportunities.
	Conversely, SMP introduces more conditionally inactive logic that TPCS can skip, while additional signal-level dependencies increase Verilator's evaluation costs.
	This is consistent with TPCS benefiting from inactive computation, although the reported speedups also use different RTL baselines and thread configurations.

\subsubsection{On Generated RTL Quality}

	TPCS achieves the shortest clock periods for the full pipeline and MDU, improving on handcrafted SV by 3.09\% and 4.74\%, with area changes of $-3.31\%$ and $+1.68\%$, respectively.
	FPU timing and area overheads are 2.58\% and 3.77\%, although PyMTL achieves a smaller area.
	The relatively modest timing differences on the full pipeline and MDU are consistent with differences in lowering: the evaluated PyMTL and SystemC flows retain more readable control structures in RTL, whereas TPCS and Chisel apply more aggressive lowering and optimization through CIRCT.
	Relative to Chisel, TPCS has shorter clock periods but slightly larger MDU and FPU areas, consistent with an area-timing tradeoff.
	Across targets, similar MDU critical paths limit variation, whereas the FPU's multiple complex arithmetic paths amplify differences in generated logic.
	These results show competitive RTL quality, with differences reflecting the generated logic structures.
	
	Disabling predicate sinking raises the full-pipeline minimum clock period to 32.7\% above the handcrafted reference; simplifying only whole predicates leaves a 10.9\% overhead.
	Including subexpression simplification instead yields the 3.09\% reduction in Table \ref{tab-direct-comp}.
	The degraded configurations contain serial Boolean dependencies from nested predicates along FMA carry-update and EXU-bypass--branch--BPU-update paths.
	Despite the shared CIRCT backend with Chisel, this ablation specifically establishes the timing benefit of predicate sinking within TPCS.

\subsection{Validation on Dataflow-Oriented Accelerators}
\label{sec-exp-asic}

	\begin{figure} [t]
		\centering
		\includegraphics[]{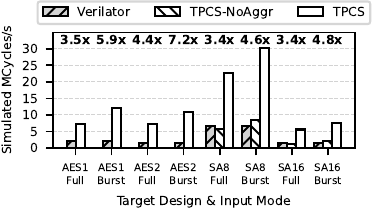}
		\caption{Simulation speed comparison on dataflow-oriented accelerator target with different input mode.}
		\label{fig-effi-aisc}
	\end{figure}

	\begin{figure} [t]
		\centering
		\includegraphics[]{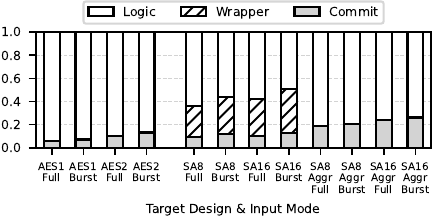}
		\caption{Overhead composition of TPCS execution on accelerator target.}
		\label{fig-compos2}
	\end{figure}

	\begin{figure} [t]
		\centering
		\subfigure[Minimum clock period.]{
			\begin{minipage}{0.45\columnwidth}
				\centering
				\includegraphics[]{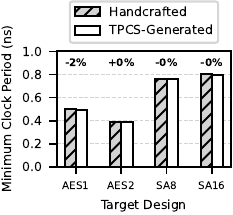}
			\end{minipage}
			\label{fig-rtlqol-time}
		}
		\subfigure[Standard-cell area.]{
			\begin{minipage}{0.45\columnwidth}
				\centering
				\includegraphics[]{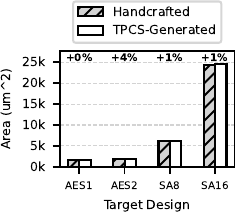}
			\end{minipage}
			\label{fig-rtlqol-area}
		}
		\caption{Timing and area performance comparison of handcrafted RTL and TPCS-generated RTL.}
		\label{fig-rtlqol}
	\end{figure}

	Fig. \ref{fig-effi-aisc} compares TPCS execution with Verilator simulation of its generated RTL on AES and systolic-array accelerators.
	Both use identical inputs and test harnesses over continuous 10-minute windows, following the Full and Burst modes defined in Section \ref{sec-exp-setup-workload}.
	TPCS achieves speedups of 3.4$\times$-4.4$\times$ in \emph{Full} mode and 4.6$\times$-7.2$\times$ in \emph{Burst} mode.
	Continuous input limits pruning opportunities in these regular datapaths, whereas input gaps increase the benefit of skipping inactive logic.
	The Full-mode results demonstrate efficient execution under sustained activity, consistent with native compilation and the absence of runtime event scheduling, although they do not isolate these benefits from pruning.

	Fig. \ref{fig-compos2} shows that transaction granularity also affects performance.
	Without array aggregation, frequent calls across small systolic-array instances introduce substantial wrapper overhead, reducing performance toward the Verilator baseline.
	Aggregation largely removes this overhead, leaving logic evaluation and register commitment as the dominant costs.
	
	Fig. \ref{fig-rtlqol} extends the RTL quality evaluation to these regular datapaths.
	Under the same OpenROAD/ASAP7 flow, TPCS-generated RTL has timing within approximately 2\% of the handcrafted references and standard-cell area overheads of approximately 0\%–4\%.
	Together with the CPU results in Section \ref{sec-exp-cmpr}, these measurements show that the TPCS compiler maintains comparable RTL quality across both control-oriented pipelines and dataflow-oriented accelerators.

\section{Discussion and Limitations}

	The experiments show that TPCS combines cycle-level agreement on the evaluated workloads with faster direct execution and competitive RTL quality.
	Simulation gains depend on input activity, transaction granularity, and parallel synchronization costs, while compiler ablation demonstrates the importance of predicate sinking for timing quality.
	These results support fast pre-RTL exploration using an executable model structurally aligned with its implementation.
	Explicit register boundaries and cycle semantics help relate backend results to modeling decisions, potentially reducing manual reimplementation and enabling earlier diagnosis of microarchitectural bottlenecks.

	TPCS nevertheless has three limitations.
	First, usability depends on infrastructure that is not yet mature, since large-scale development requires stronger modeling diagnostics, debugging support, and reusable libraries.
	The native C++ representation also offers opportunities for AI-driven development, but reliable use requires dedicated validation and debugging infrastructure in the future.

	Second, hardware coverage remains to be extended beyond the evaluated cycle-synchronous designs.
	The current TPCS semantics provide a sufficient foundation for synchronous composition, but broader applicability requires standardized clock-domain-crossing (CDC) components, explicitly defined asynchronous primitives, and gated clocks.
	Although faithfully capturing physical metastability at this abstraction level is impractical, configurable-delay models may support asynchronous-synchronous behavioral equivalence analysis under explicit timing assumptions.

	Third, the compiler currently emphasizes combinational timing optimization, with limited support for automatic pipelining and systematic area-timing exploration.
	A future pipelining extension could let users specify the latency from block activation to output transactions, reducing practical development efforts.
	Closer integration with gate-level and physical-design optimization could further improve RTL quality.

\section{Conclusion}

	This paper presents Transaction-Pruned Cycle Simulation (TPCS), a modeling and simulation methodology for cycle-synchronous hardware.
	TPCS maps hardware control flow onto native C++ execution to prune inactive computation without explicit event scheduling, while a dedicated compiler generates structurally corresponding RTL.
	Experiments show up to 42$\times$ speedup over parallel Verilator on SMP systems, 4.77$\times$ over handcrafted SV simulation on an isolated CPU pipeline, and 3.4$\times$–7.2$\times$ on accelerators.
	The evaluated designs retain cycle-level waveform agreement and achieve RTL timing and area comparable to handcrafted implementations.
	These results support a unified representation for fast pre-RTL exploration, validation, and hardware implementation.
	
	Future work will improve tooling and AI-assisted infrastructure, broaden hardware coverage through standardized CDC components and asynchronous primitives, and extend the compiler with latency-aware automatic pipelining, gate-level optimization, and systematic area–timing exploration.

\section*{Acknowledgment}

	This work is supported by a series of significant research projects, including the National Natural Science Foundation of China (Grant No. U25A6023), "New Intelligent Processing Architecture and System Based on RISC-V Artificial Intelligence Instruction Set Extension", the National Key R\&D Program of China (Grant No. 2023YFB4503901), "Research on RISC-V Based Software Ecosystem Technologies for Data Centers", and the Jinan "New Universities 20" Initiative, especially the "Autonomous Controllable High-Performance RISC-V Chip Innovation Team" (Grant No. 202333041).

\bibliographystyle{ieeetr}
\bibliography{citation}

{
\setlength{\parskip}{0pt}
\setlength{\itemsep}{0pt}
	
\vspace{-1.0in}
\begin{IEEEbiography}[{{\includegraphics[width=0.8in]{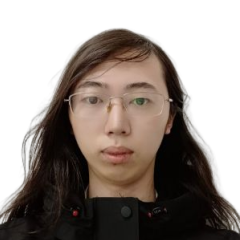}\vspace{0.4in}}}]{Chengzhen Meng}
	is currently a Ph.D. student in software engineering at the School of Software, Shandong University. His research interests include novel computer architecture design and optimization with hardware–software co-design and design automation.
\end{IEEEbiography}
\vspace{-1.5in}
\begin{IEEEbiography}[{{\includegraphics[width=0.8in]{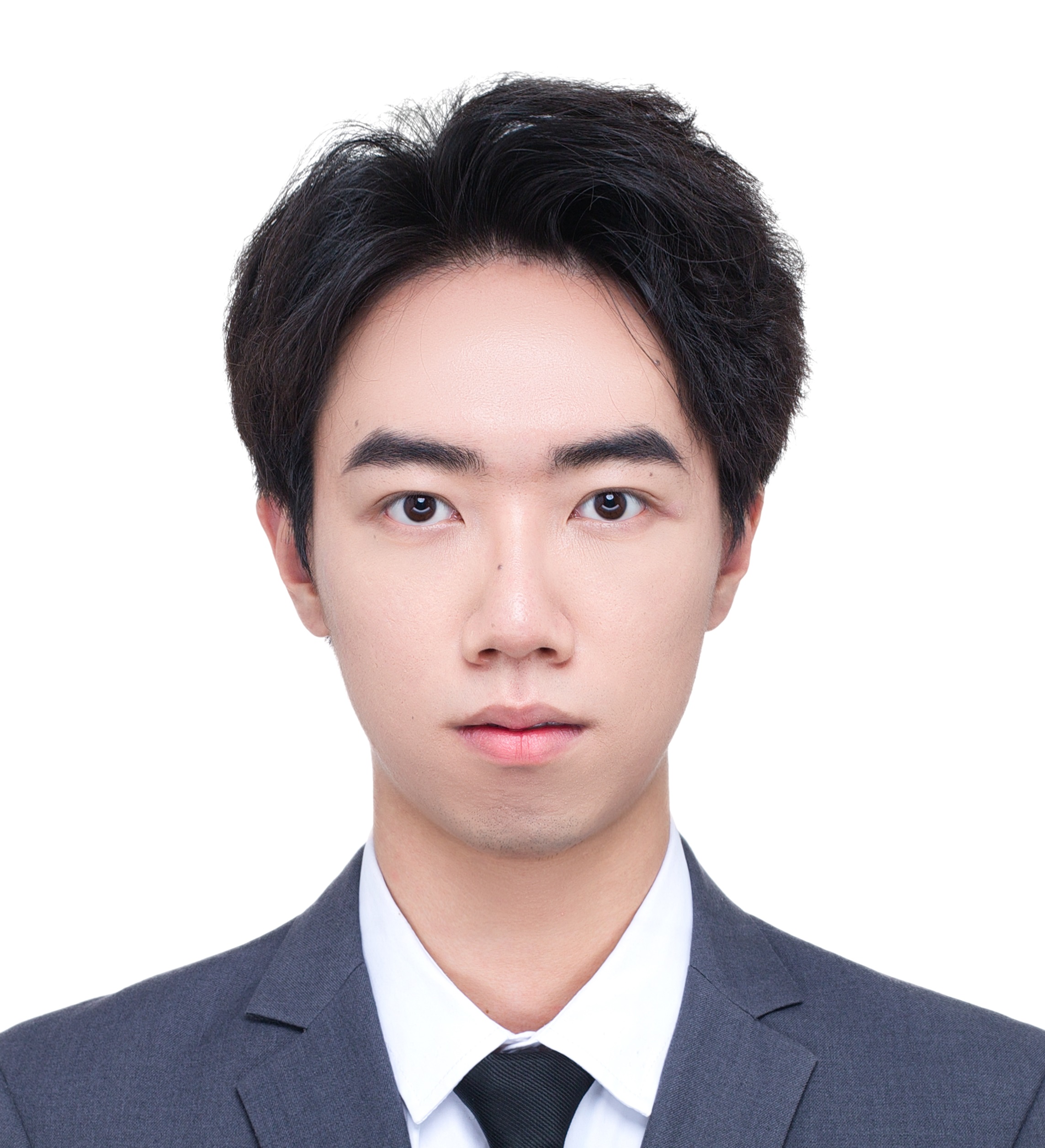}\vspace{0.4in}}}]{Xiaoyu Li}
	is currently a Master’s student in Integrated Circuit Engineering at the School of Integrated Circuits, Shandong University. His research interests include digital circuit design and computer architecture for high-performance platforms.
\end{IEEEbiography}
\vspace{-1.5in}
\begin{IEEEbiography}[{{\includegraphics[width=0.8in]{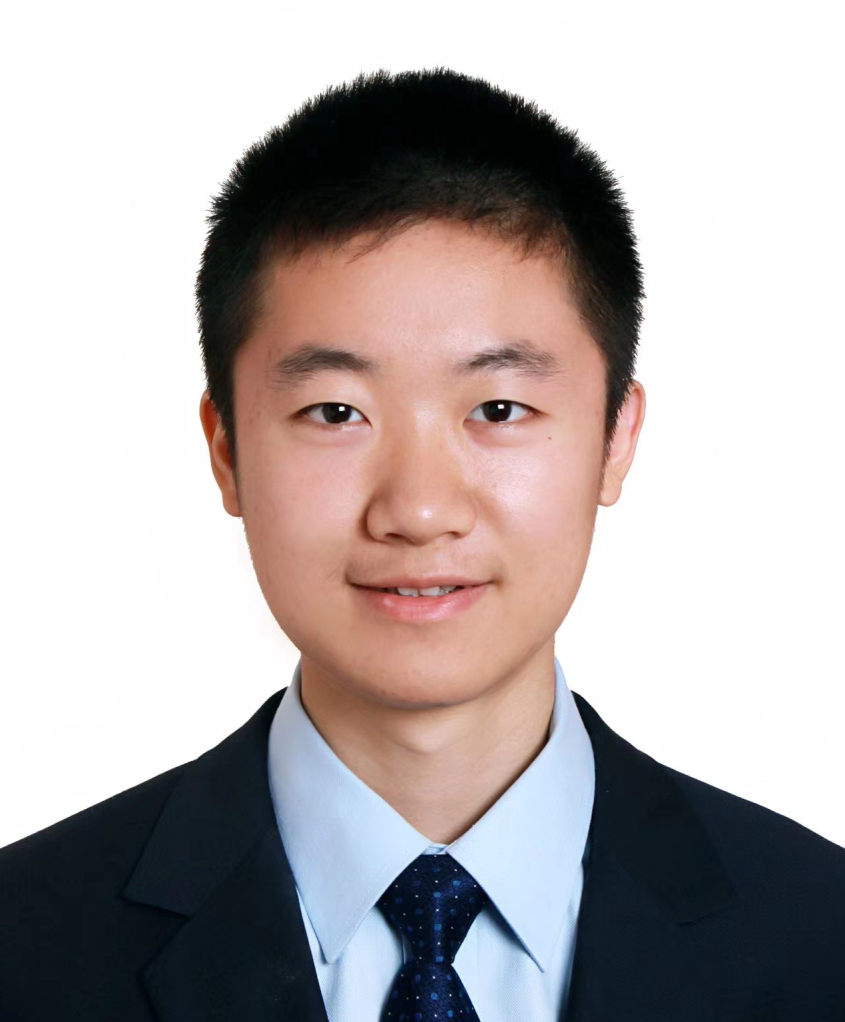}\vspace{0.4in}}}]{Chen Bai}
	received the Ph.D. degree in computer science and engineering from The Chinese University of Hong Kong in 2024. He is currently a Postdoctoral Fellow with the Department of Electronic and Computer Engineering, The Hong Kong University of Science and Technology. He will join Fudan University as a tenure-track assistant professor. His research interests include computer architecture and electronic design automation. He received the William J. McCalla Best Paper Award at ICCAD 2021, the Best Paper Award at ASPLOS 2026, and a Best Paper Award Nomination at ISPD 2024.
\end{IEEEbiography}
\vspace{-1.2in}
\begin{IEEEbiography}[{{\includegraphics[width=0.8in]{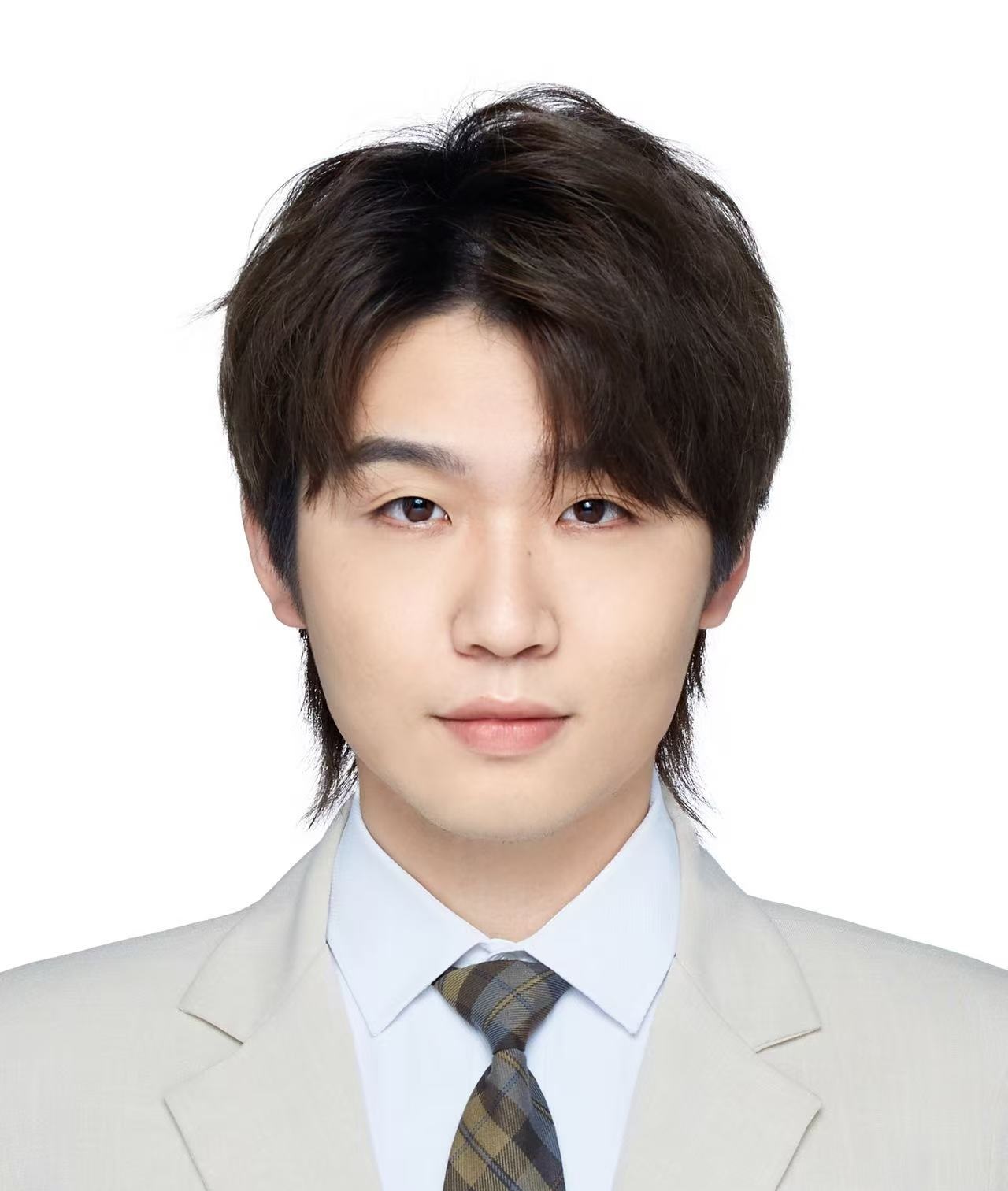}\vspace{0.4in}}}]{Zhenyu Zhang}
	is currently a Ph.D. student at the School of Integrated Circuits, Shandong University. His research lies at the intersection of machine learning system architecture and AI infrastructure. Specifically, he focus on AI compilers (such as MLIR and TVM), hardware acceleration across GPUs and TPUs, and the optimization of large language model (LLM) operators.
\end{IEEEbiography}
\vspace{-1.5in}
\begin{IEEEbiography}[{{\includegraphics[width=0.8in]{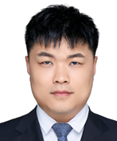}\vspace{0.4in}}}]{Hongjun Dai (Member, IEEE)}
	received the BE and PhD degrees in computer science from Zhejiang University, China, in 2002 and 2007, respectively. He is a professor of software engineering with Shandong University, China. He has authored or coauthored more than 30 peer-reviewed journal and conference papers. His research interests include novel computer architecture optimization and the framework of distributed systems such as IoT and cloud.
\end{IEEEbiography}
	
}

\end{document}